\documentclass[a4paper,10pt]{article}
\usepackage[dvips]{graphicx}
\usepackage{amsfonts,amsmath,amssymb}

\begin{document}
\title{Vacuum Polarization and Casimir Energy of a Dirac Field Induced by a Scalar Potential in One Spatial Dimension}

\author{S.S. Gousheh\footnote{Electronic address: ss-gousheh@sbu.ac.ir}, S.S. Mousavi and L. Shahkarami\footnote{Electronic address: l\_shahkarami@sbu.ac.ir}\\
    Department of Physics, Shahid Beheshti University G.C., Evin, Tehran
19839, Iran}
\maketitle
\begin{abstract}
We investigate the vacuum polarization and the Casimir energy of a Dirac field coupled to a scalar potential in one spatial dimension. 
Both of these effects have a common cause which is the distortion of the spectrum due to the coupling with the background field. 
Choosing the potential to be a
symmetrical square-well, the problem becomes exactly solvable and we can
find the whole spectrum of the system, analytically. 
We show that the total number of states and
the total density remain unchanged as compared with the free case, as one expects.
Furthermore, since the positive- and negative-energy eigenstates of the fermion are fermion-number 
conjugates of each other and there is no zero-energy bound state, the total density and 
the total number of negative and positive states remain unchanged, separately.
Therefore, the vacuum polarization in this model is zero for any
choice of the parameters of the potential. 
It is important to note that although the vacuum polarization is zero due to the symmetries of the model, 
the Casimir energy of the system is not zero in general. 
In the graph of the Casimir energy as a function of the depth of the well there is a maximum approximately 
when the bound energy levels change direction and move back towards their continuum of origin. 
The Casimir energy for a fixed value of the depth is a linear function of the width and is always positive.
Moreover, the Casimir energy density (the energy density of all
the negative-energy states) and the energy density of all the
positive-energy states are
exactly the mirror images of each other. 
Finally, computing the total energy of a valence fermion present in the lowest fermionic bound state, 
taking into account the Casimir energy, we find that the lowest bound state is almost always unstable for the scalar potential.
\end{abstract}
\section{Introduction}
During the last forty years the concept of the vacuum polarization (VP) of fermions due to their interactions with the other fields has attracted much interest.
It has been studied extensively in many branches of physics such as
particle physics \cite{jackiw,goldstone,mackenzie1,particle1,heavy1,dr,heavy4,heavy6,particle4,me}, condensed-matter physics \cite{cond}, polymer physics \cite{poly}, atomic physics \cite{atom}, and cosmology \cite{cosm}. 
Most of the authors have considered the coupling of scalar and pseudoscalar fields to fermions and obtained many interesting and unexpected results. 
In this category of problems, solitary waves and solitons have been chosen extensively as background fields. 
An interesting phenomenon, when a fermion interacts with a soliton, is the fractionalization of the fermion number of the solitonic states. 
This was first pointed out by Jackiw and Rebbi \cite{jackiw}. 
They showed that in the fermion-soliton models in which the soliton is in the form of a
prescribed scalar field and the system possesses the charge conjugation symmetry, the existence of a zero-energy fermionic mode, implies that the soliton is a degenerate doublet with fermion number $\pm 1/2$. 
This interesting result has motivated much of the works in VP for many different physical systems in the literature. 
\par    
Two important tools for studying the vacuum polarization were invented in the early 80s. 
The first one, called the adiabatic method, was introduced by Goldstone and Wilczek \cite{goldstone}. 
In this method the nontrivial configuration of the external background field is considered to form continuously and adiabatically (slowly) from  the trivial configuration. 
Using their adiabatic method for coupled fermion-soliton systems which do not possess charge conjugation symmetry, they concluded that for such systems the fermion number of the soliton could be any real value, not just $\pm1/2$. 
Later on the limitation of this method, i.e.\;the requirement of adiabaticity for the external field, was lifted by MacKenzie and Wilczek \cite{mackenzie1}. 
In this modified method one computes the energy spectrum of the fermion as the prescribed field is formed. 
They applied this method to compute the vacuum polarization of a Dirac field, induced by an infinitely sharp soliton as an example \cite{mackenzie2}. 
Using their method they concluded that sharply varying solitons do not carry any fermionic charge. 
According to their method, there are in general two contributions to VP. 
First is the adiabatic contribution which comes from the change in the total number of
energy levels in the Dirac sea, caused by the presence of the disturbance. 
The second contribution, which is called the nonadiabatic contribution, is due to the bound
states with negative energy.
In the expansion of the Fermi field operator in terms of the fermionic states in the presence of the disturbance, they chose the coefficients of the bound states to be  always the absorption operator.
With this choice just the adiabatic contribution is included in VP and one has to add the other contribution by hand.
However, we choose the coefficients of the bound states with positive and negative energy to be the annihilation and creation operator, respectively, as one does for the continuum states.
This appropriate choice leads to a complete formula for VP, in which both contributions are automatically included. 
With the aid of the method of MacKenzie and Wilczek, Gousheh and Mobilia \cite{dr}
computed the vacuum polarization by solitons for an exactly solvable model. 
Their model is a Fermi field chirally coupled to a pseudoscalar field with a simple form similar to the kink or the soliton of sine-Gordon model. 
By the use of this exactly solvable model, they concluded that solitons in general could polarize the vacuum, and only the infinitely sharp solitons never polarize the vacuum.
\par 
Another manifestation of the distortion of the fermionic spectrum is the Casimir effect. The standard Casimir effect was first proposed theoretically by Dutch physicist Hendrik Casimir \cite{casimir1,casimir2} in 1948.
He predicted the existence of a net attractive force between two grounded infinite parallel metallic plates in a vacuum without any external electromagnetic field. 
The plates change the zero energies of the quantized fields and give rise to forces between them. 
In 1958 Marcus Sparnaay \cite{sparnaay} conducted the first experimental attempt to observe this phenomenon for two parallel metallic plates. 
However, the results had a very poor accuracy since two parallel plates would require accurate alignment to ensure they are parallel.
In 1997 Steve K. Lamoreaux \cite{Lamorea1} opened the door to precise measurements of the Casimir force, using a plate and a metallic spherical shell. 
Since then, many different experiments have been performed to measure the Casimir forces for various geometries \cite{exper}. 
Although many theorists refer to the Casimir force as a good evidence of the reality of quantum zero-point fluctuations, some authors \cite{vaccas4} believe that this effect gives no support for the reality of the vacuum energy.
The Casimir effect can be observed when the presence of some nontrivial boundary condition or background field (e.g.\;a soliton) changes the vacuum energy of a quantum field. 
Also, the value of the Casimir energy depends on the number of spatial dimensions, the type of fields, type of topology, and geometry. 
Since the Casimir's work
many papers have been written on the Casimir energy for different geometries 
\cite{casimir1,boyer,deraad,geometry} using 
many different techniques \cite{techniques} to remove the divergences.
There are many recent works in which they study the practical applications of the Casimir effect. 
The Casimir forces which are normally neglected in macro systems have to be considered for Micro and Nano Electromechanical Systems (MEMS and NEMS) \cite{mems}.
\par
As mentioned above, the zero-point energy can also be affected by the presence of
nontrivial background fields such as solitons. 
Several authors have used various methods to compute the Casimir energy caused by the presence of solitons and specially to compute the corrections to the soliton mass, such as the dimensional regularization, zeta function analytic continuation and scattering phase shift method
\cite{heavy1,heavy4,solmass,la,al,al2}.
Also, the Casimir effect appears in supersymmetric models to investigate the validity of the BPS saturation by supersymmetric solitons \cite{super}.
Choosing a soliton as the background field for investigating the Casimir energy in the presence of the nontrivial background fields, usually makes the problem analytically
unsolvable 
and the problems of this kind which are exactly solvable are very rare \cite{la,al}. 
The use of numerical methods might hide some important physical aspects of these problems. 
Therefore, choosing simple background fields which give rise to exactly solvable
problems could clarify some of the physical aspects of the Casimir effect and also the vacuum polarization. 
For this purpose we choose a simple model in which a Dirac field is coupled to a scalar
potential in ($1+1$) dimensions. 
The simple functional form chosen for the potential which is a symmetrical square-well makes the problem exactly solvable. 
Our model has the symmetries C, P and T, separately. 
The charge conjugation operator in this model has the property of taking a solution with positive energy $E$ into the one with the negative energy $-E$.
Also, we observe that for the scalar potential chosen here there is no zero-energy bound state.
This fact and the existence of the symmetry between the negative- and 
positive-energy eigenstates lead to some interesting results in VP and the Casimir energy of this model which we will state and study throughout this paper.
\par
It is worth mentioning that it is traditional to discuss of the Casimir effect when
a dynamical field interacts with a nontrivial boundary condition or a nontrivial 
``topological'' background field.
However, many authors generalize the Casimir effect to the situation in which a 
nontrivial ``nontopological'' background field, like the one we consider in this paper, 
is present and changes the vacuum energy of the dynamical field (see for 
example \cite{nontopol}).
\par
In section II we introduce the model and present the complete spectrum of the Fermi field in the presence of the potential.
In section III we compute the vacuum charge of this system by the use of the method proposed by MacKenzie and Wilczek \cite{mackenzie1}.
In this section we show that not only the change in the total number of levels due to the potential well is always zero, but also the total number of levels with negative and positive energy, separately, is exactly the same as the case of free Dirac field. 
Therefore, the two contributions for VP in our model always cancel each other, i.e.\;the scalar potential coupled to the Fermi field never polarizes the vacuum.
In section IV we calculate and depict the densities of bound states and the difference between the spatial densities of the wave functions in the presence and absence of the potential for the negative continuum and also the positive continuum, for comparison. 
By the use of these investigations we conclude that the spatial density of the spectrum
remains uniform in the presence of the square-well. 
Furthermore, for this problem the total density of states for states with $E<0$ and $E>0$ are separately unchanged from the free case. 
This also happens due to the symmetry in the energy spectrum of the fermion.
In section V we thoroughly explore the behavior of the Casimir energy and the energy densities for our model. 
First, we present a general expression for the Casimir energy of a Fermi field in the presence of a general disturbance (a background field or nontrivial boundary condition), by subtracting the zero-point energies in the presence and absence of the disturbance
\cite{la}. 
Then, we obtain an exact expression for the Casimir energy of our model and investigate the behavior and properties of this energy as we vary the depth and width of the potential well. 
We conclude the interesting result that although VP is always zero for our model, its Casimir energy is in general nonzero.
Then, we explore the behavior of the distortion of the energy densities in the continua 
and see that the total negative- and positive-energy densities are exactly the mirror images of each other. 
In section VI we compute the total energy for a system consisting of a valence fermion in the first bound state and conclude that this bound state is unstable for most choices of the parameters of the scalar potential.
In section VII we summarize the results and state some conclusions.
\section{The spectrum of a Dirac particle in a one-dimensional square-well potential}
The Dirac equation in a one-dimensional scalar potential can be written as
\begin{equation}\label{e1}\vspace{.2cm}
 \left[i \gamma^\mu \partial_\mu
 -( m_0 +V(x))\right]\psi \left(x,t\right)=0 ,\hspace{.4cm}  \mu= 0,1.
\end{equation}
 Our first task is to solve this equation. 
We choose the following representation for the Dirac matrices: $\gamma^0=\sigma_1$  and  $\gamma^1=i\sigma_3$, in which the Dirac equation becomes
\begin{equation}\label{e2}\vspace{.2cm}
 \left[i \sigma_1 \partial_t-\sigma_3\partial_x
 -( m_0 +V(x))\right]\psi \left(x,t\right)=0.
\end{equation}
In ($1+1$) dimensions the Fermi field can be written as
$\psi=\dbinom{\psi_1} {\psi_2}$. We define
\begin{equation}\label{e3}
\xi(x,t)=e^{-iEt}\dbinom{\xi_1(x)}{\xi_2(x)}=\dbinom{\psi_1+i\psi_2}
{\psi_1-i\psi_2}.
\end{equation}
The equations of motion in terms of $\xi_1(x)$ and $\xi_2(x)$ are as follows
\begin{align}\label{e4}
 \bigg \{\partial_x^2-\frac{V'(x)}{m_0+V(x)}\partial_x+
 E^{2}-\left(m_0+V(x)\right)^2\mp\frac{iEV'(x)}{m_0+V(x)}\bigg\}\xi_{1,2}(x) =0.
\end{align}
We choose the functional form of $V(x)$ to be a symmetrical square-well potential with depth $-V_0\leq0$ and width $2a$, where the values for $a$ are in units of the inverse mass of the fermion, $m_0$. 
The simple functional form chosen for $V(x)$ renders the problem exactly solvable. 
The potential well acts as a background field that modifies the Dirac spectrum. 
In particular the number of states in the positive and negative continua changes and bound states appear.
As is well-known, these changes are the sources for the vacuum
polarization and Casimir energy. 
To investigate these interesting phenomena, we first need to study the complete spectrum of the fermion in the presence of the background field. 
Since the Hamiltonian is invariant under parity, the eigenfunctions of the
Hamiltonian can be chosen to be eigenfunctions of the parity
operator as well. 
In the first representation the parity operation
is given by $P\psi \left(x,t\right)=\sigma_1\psi \left(-x,t\right)$.
In the representation of Eq.\,(\ref{e3})
 $(\gamma^0=\sigma_2$ and $\gamma^1=i\sigma_1)$ it becomes $P\xi \left(x,t\right)=-\sigma_2\xi\left(-x,t\right)$.
The resulting bound states are of the following form
\begin{equation}\label{e5}\vspace{.2cm}
\xi_{\text{b}}(x)=\begin{cases}\vspace{.3cm}
N\dbinom{1}{-\frac{(\lambda+iE)}{m_0}}\text{e}^{\lambda(x+a)},& x
\leqslant -a ,\\\vspace{.3cm}
\dbinom{c\,{\text{e}^{i\mu x}}+d\,{\text{e}^{-i\mu x}}}{c\,f_{+}{\text{e}^{i\mu x}}+d\,f_{-}{\text{e}^{-i\mu x}}}, & -a\leqslant x\leqslant a,\\
b\dbinom{1}{\frac{(\lambda-iE)}{m_0}}\text{e}^{-\lambda(x-a)}, &
x \geqslant a ,\\
\end{cases}
\end{equation}
where $\mu=\sqrt{E^2-(m_0-V_0)^2}$, $\lambda=\sqrt{m_0^2-E^2}$,
$f_{+}=\frac{-i(E+\mu)}{(m_0-V_{0})}$,
$f_{-}=\frac{-i(E-\mu)}{(m_0-V_{0})}$. From the continuity of $\xi_\text{b}$
at $x=\pm a$ and the normalization condition we obtain the coefficients
which are shown in the Appendix A. 
The bound state energies are solutions to the following equation
\begin{equation}\label{e6}\vspace{.2cm}
 \frac{\mu\lambda}{m_0V_0}\cot(2a\mu)+\frac{\lambda^2}{m_0V_0}=1.
\end{equation}
After some calculations, this equation can be written in the form
$\frac{E V_0}{\lambda\mu} \sin(2a\mu)=\pm1$ in which $\pm$ signs
refer to the parity eigenvalues.
\par
\begin{center}
\begin{figure}[th] \hspace{1.2cm}\includegraphics[width=13.cm]{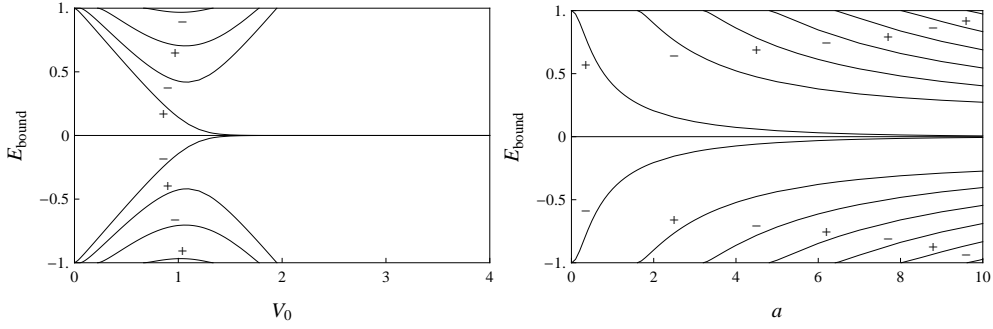}\caption{\label{fig.2} \small
   Left graph: The energies of the bound states as a function of $V_0$ at $a=5$.
Right graph: The energies of the bound states as a function of $a$ at $V_0=1.2$.
   The values obtained are the solutions to the transcendental Eq.\,(\ref{e6}).
   The parity of each of the bound states is indicated on the graphs by $\pm$ signs.
Note that no bound state crosses $E=0$.}
\end{figure}
\end{center}

 
Throughout this paper we demonstrate most of the results for  $0\leqslant V_0\leqslant4m_0$. Also, we rescale the quantities of the
system with respect to the fermion mass ($m_0$) and work with the
dimensionless quantities, for simplicity. 
We illustrate the results of bound state energies obtained using Eq.\,(\ref{e6}) in 
Fig.\,\ref{fig.2}. 
The parities of the bound states are indicated on the graphs by $\pm$
signs.
Our Lagrangian has all the symmetries C, P and T, separately.
The charge conjugation operator in the first representation ($\gamma^0=\sigma_1$  and  $\gamma^1=i\sigma_3$) is $\sigma_3$ which relates the positive- and 
negative-energy eigenstates as $\sigma_3 \psi^{*}_{E}=\psi_{-E}$.
Therefore, this operator has the property of taking a solution with eigenvalue $E$ into the one with the eigenvalue $-E$. 
This symmetry is obvious from the graphs of the bound energy levels.

\par
The explicit expression for the eigenfunctions in the continua is as follows
($\pm$ signs refer to the parity eigenvalues)
\begin{equation}\label{e7}\vspace{.4cm}
\xi_p^{\pm}(x)=\begin{cases}\vspace{.3cm}
h\dbinom{1}{-i\frac{(p+E)}{m_0}}\text{e}^{ip(x+a)}+k\dbinom
{1}{i\frac{(p-E)}{m_0}}\text{e}^{-ip(x+a)},& \hspace{.2cm} x
\leqslant -a ,\\\vspace{.3cm}
N_{\pm}\dbinom{1}{i\frac{(V_0-m_0)}{(E-\mu)}}\text{e}^{i\mu
x}\pm N_{\pm}\dbinom { \frac{(m_0-V_0)}{(E-\mu)}}{-
i}\text{e}^{-i\mu x},
 &\hspace{.2cm} -a\leqslant x\leqslant a,\\
\pm k\dbinom{ \frac{(E-p)}{m_0}}{-i}\text{e}^{ip(x-a)}\pm h\dbinom {
\frac{(p+E)}{m_0}}{- i}\text{e}^{-ip(x-a)}, &\hspace{.2cm} x
\geqslant a ,
\end{cases}
\end{equation}
where $p^2=E^2-m_0^2$. The explicit expressions for all the parameters
in this equation are given in the Appendix A. 
We have normalized the wave functions by the following relation
\begin{equation}\label{e8}\vspace{.2cm}
\sum_{j=\pm} \int_{-\infty}^{+\infty}
\xi_{p}^{j\dag}(x)\xi_{p'}^j(x)\text{d}x=2\pi\delta(p-p^\prime).
\end{equation}
Moreover, the states of the free Dirac particle are given by
\begin{equation}\label{e9}\vspace{.2cm}
 \xi^{\text{free}}_k(x)=\dfrac{m_0}{\sqrt{2E(E+k)}}\dbinom{1}{-i\frac{(k+E)}{m_0}}\text{e}^{ikx}.
\end{equation}
\section{The vacuum charge}
In this section we explore the vacuum polarization of the Fermi field in the presence of the potential well. 
The potential well acts as a background field that generates a distortion in the whole spectrum of the fermion. 
First, we show that the normalized vacuum charge is related to the difference between the number of negative-energy levels in the presence
 and absence of the disturbance, or equally the difference between the number of positive-energy levels with an overall minus sign. 
The eigenfunctions of the free Dirac Hamiltonian form a complete set. 
We assume that the set of solutions in the presence of the disturbance is also complete. 
Hence, the Fermi field operator can be expanded in terms of either of these two complete orthonormal  bases as follows
\begin{align}\label{e10}\vspace{.2cm}
\Psi(x)&=\int_{-\infty}^{+\infty}\frac{\text{d}k}{2\pi}[b_ku_k(x)+d^{\dag}_kv_k(x)]\\\label{e10a}
&=\int_0^{+\infty}\frac{\text{d}p}{2\pi}\sum_{j=\pm}[a^j_p\mu_p^j(x)+c_p^{j\dag}\nu_p^j(x)]
+\sum_i[e_i\chi_{1\text{b}_i}(x)+f^{\dag}_i\chi_{2\text{b}_i}(x)],
\end{align}
where we have denoted  $\xi_k^{\text{free}}(x)$ by $u_k(x)$  and $v_k(x)$ for the states with $E>0$ and $E<0$, respectively. 
Similarly $\xi_p(x)$ which are the wave functions for the continua in the presence of the
well, are denoted by $\mu_p(x)$ in the positive continuum and by $\nu_p(x)$ in the negative continuum. 
Also, the bound state wave functions, $\xi_{\text{b}}(x)$, are separated into 
positive-energy ones denoted by $\chi_{1\text{b}_i}(x)$ and negative-energy ones denoted by $\chi_{2\text{b}_i}(x)$. 
We choose the annihilation (creation) operator for the bound states with positive
(negative) energy, as we do for the continuum states. 
We shall see one of the advantages of this choice when we compute the vacuum
polarization and Casimir energy. 
Imposing the usual  anticommutation relations between $\Psi$  and  $\Psi^{\dag}$, or the resulting anticommutation relations between the creation and annihilation
operators, the number operator in the free case becomes
\begin{equation}\label{e11}\vspace{.2cm}
 N=b^{\dag}_kb_k-d^{\dag}_kd_k.
\end{equation}
Using orthonormality of both sets of eigenfunctions, one can express  $b$ and $d$
operators in terms of $e$,  $f$, $a$ and $c$. Then, the number operator becomes
\begin{equation}\label{e12}\vspace{.2cm}
 N=e^{\dag}_ie_i+a^{j\dag}_pa_p^j-c^{j\dag}_pc_p^j+\left\langle \chi_{2\text{b}_{i}}|\chi_{2\text{b}_i}\right\rangle
 +\left\langle \nu^{j}_p|\nu_p^j\right\rangle-\left\langle v_k|v_k\right\rangle.
\end{equation}
The last two terms are the difference between the total number of
levels in the Dirac sea in the presence and absence of the
disturbance and the term $\left\langle \chi_{2\text{b}_{i}}|\chi_{2\text{b}_i}\right\rangle$ is the number of the negative bound states.
The changes in the number of levels in the negative continuum ($Q_{\text{sea}}$) and positive continuum ($Q_{\text{sky}}=\langle \mu^{j}_p|\mu_p^j\rangle-\left\langle u_k|u_k\right\rangle$) are given by
\begin{align}\label{e13}\vspace{.2cm}
Q_{\begin{smallmatrix} \text{sky}\\ \text{sea}
\end{smallmatrix}}&=\sum_p\Big\langle
\xi_{p_{\begin{smallmatrix} \text{sky}\\ \text{sea}
\end{smallmatrix}}}{\Big |}\xi_{p_{\begin{smallmatrix} \text{sky}\\ \text{sea}
\end{smallmatrix}}}\Big\rangle-\sum_k\big\langle
\xi_{k}^{\text{free}}\big|\xi_{k}^{\text{free}}\big\rangle\nonumber\\&=\sum_{j=\pm}\int_{0}^{+\infty}\frac{\text{d}p}{2\pi}\int_{-\infty}^{+\infty}\text{d}x\;
\xi^{j\dag}_{p_{\begin{smallmatrix} \text{sky}\\ \text{sea}
\end{smallmatrix}}}\xi^{j}_{p_{\begin{smallmatrix} \text{sky}\\ \text{sea}
\end{smallmatrix}}}
-\int_{-\infty}^{+\infty}\frac{\text{d}k}{2\pi}\int_{-\infty}^{+\infty}\text{d}x\;\xi^{\text{free}\dag}_{k}\xi^{\text{free}}_{k}.
\end{align}
\begin{center}
\begin{figure}[th] \hspace{3.5cm}\includegraphics[width=8.cm]{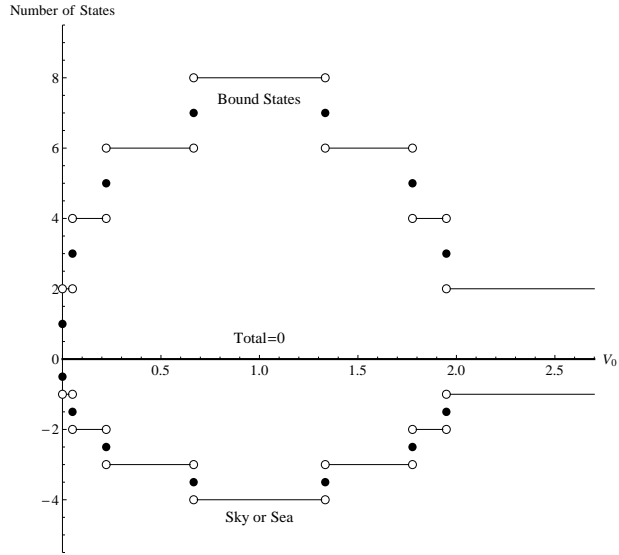}\caption{\label{fig.3} \small
The change in the number of energy levels in the negative continuum
($Q_{\text{sea}}$), positive continuum ($Q_{\text{sky}}$), and the
number of bound states, all due to the presence of the background field, and the total number of levels as a function
of $V_0$ for $a=5$. 
We see that the trend for both $Q_{\text{sea}}$
and $Q_{\text{sky}}$ is exactly the same, as an expected result of
the particle conjugation symmetry of the Lagrangian. 
Note that the number of states associated with each of
the continua is actually the difference between the number of levels
in that continuum in the presence and absence of the potential
well.}
\end{figure}
\end{center}
Our prescription for subtracting the two divergent integrals in Eq.\,(\ref{e13}) and the other similar relations to follow is to combine the integrals and subtract the integrands with the same values of $p$ and $k$. 
The integral over the spatial variable in Eq.\,(\ref{e13}) can be performed analytically.
However, the leftover integral over $p$ cannot be performed analytically and we calculate it numerically and show the results in Fig.\,\ref{fig.3}, along with the graphical representation of the number of bound states and the total number of levels as a function of $V_0$ for $a=5$. 
From this figure we observe that the general trend for $Q_{\text{sea}}$ is constant superimposed with jumps of minus one whenever a bound state peels off (separates) from the sea (at $E=-m_0$). 
At the points where the separation occurs, the jump is $-1/2$. 
There is a jump of plus one whenever a bound state joins the sea and at the points where the joining occurs, the jump is $+1/2$. 
As can be seen from the figure, the deficiency in the number of continuum states with  negative energy is minus one for $V_0>2$ and this happens because for these values of $V_0$ there are only two bound states with the energies approaching $E=0$.
As expected from the symmetry of the system for the negative- and positive-energy levels, there is an exactly similar trend for $Q_{\text{sky}}$. 
In this figure we have also plotted the total number of levels as compared with the free case, i.e.\;the sum of the changes in the number of levels in the sea and the sky and the number of bound states. 
Note that the change in the total number of levels as compared with the free case is zero for all values of $V_0$. 
If we denote the number of bound states by $N_{\text{b}}$, we can express this conclusion by the following equation
\begin{equation}\label{e14}\vspace{.2cm}
Q_{\text{sea}}+Q_{\text{sky}}+N_{\text{b}}=0.
\end{equation}
We can define $N_{\text{b}}=n^>+n^<$, where $n^>$ and $n^<$ denote
the number of bound states with positive and negative energy, respectively.
Since the fermionic vacuum is defined as the state in which all of the negative-energy states are filled and the positive ones are empty, the vacuum polarization
(VP) is simply given by the following equation
\begin{equation}\label{e13a}\vspace{.2cm}
\text{VP}=Q_{\text{sea}}+n^<=-(Q_{\text{sky}}+n^>).
\end{equation}
The last equality is obtained by the use of Eq.\,(\ref{e14}).
One can obtain this formula for VP with the aid of the relation we obtained for the number operator in Eq.\,(\ref{e12}).
To this end, one
should compute the vacuum expectation value of the number operator
in Eq.\,(\ref{e12}). 
Therefore, only three last terms contribute in
the VP. 
As we mentioned before, the term 
$\left\langle\chi_{2\text{b}_{i}}|\chi_{2\text{b}_i}\right\rangle$ is the number
of bound states with negative energy ($n^<$). 
The automatically
appearance of this term in the definition of the number operator is
the advantage of including both annihilation and creation operators
for the bound states in the expansion of the Fermi field in the
presence of the disturbance. 
By using some changes in the
computation of the number operator, the last three terms in
Eq.\,(\ref{e12}) could be replaced by $-\left\langle
\chi_{1\text{b}_{i}}|\chi_{1\text{b}_i}\right\rangle
 -\langle \mu^{j}_p|\mu_p^j\rangle+\left\langle u_k|u_k\right\rangle$, 
which corresponds to the last equality in the Eq.\,(\ref{e13a}).
\par 
Now, we compute VP of the fermion in the presence of the potential well by using Eqs.\,(\ref{e13},\ref{e13a}) and the information contained in 
Fig.\,\ref{fig.2} about $n^<$. 
We conclude that as expected the vacuum polarization is zero,
regardless of the values of the parameters, i.e.\;$a$ and $V_0$. 
Note that this result is due to the charge conjugation symmetry of the system. 
Since this symmetry is present in our problem and there is no state with $E=0$, the change in the total number of levels with negative energy is always zero. 
However, for systems which do not possess this symmetry, the presence of the potential could in general polarize the vacuum 
(e.g.\;\cite{goldstone,mackenzie1,dr,me,dehghan}).
On the other hand, in the case of the Jackiw-Rebbi model \cite{jackiw} although it possesses the charge conjugation symmetry,
the vacuum polarization is not zero.
In fact a zero-energy fermionic mode which is always present is the origin of the 
nonzero VP in this model \cite{farid}.
\section{Densities of the solutions}
\begin{center}
\begin{figure}[th]\hspace{3.5cm} \includegraphics[width=8.cm]{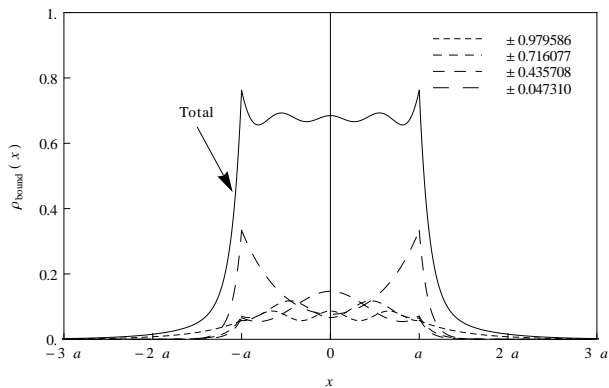}\caption{\label{fig.4} \small
The bound state densities for the potential well with $a=5$ and $V_0=1.2$. 
The individual densities are depicted with various dashed lines, as indicated on the graph,
  and their total sum with the solid line.}
\end{figure}
\end{center}

The change in the charge densities for the Dirac sea and sky is
given by
\begin{align}\label{e15}\vspace{.2cm}
\rho_{\begin{smallmatrix} \text{sky}\\ \text{sea}
\end{smallmatrix}}(x)=\sum_{j=\pm}\int_{0}^{+\infty}\frac{\text{d}p}{2\pi}\; \xi^{j\dag}_{p_{\begin{smallmatrix} \text{sky}\\ \text{sea}
\end{smallmatrix}}}(x)\;\xi^{j}_{p_{\begin{smallmatrix} \text{sky}\\ \text{sea}
\end{smallmatrix}}}(x)-\int_{-\infty}^{+\infty}\frac{\text{d}k}{2\pi}\;\xi^{\text{free}\dag}_{k}(x)\;\xi^{\text{free}}_{k}(x),
\end{align}
where $\pm$ signs refer to the positive and negative parity. 
Also, the bound state densities are simply
\begin{equation}\label{e16}\vspace{.2cm}
\rho_{\text{bound}}(x)=\sum_i\xi_{\text{b},i}^{\dag}(x)\xi_{\text{b},i}(x).
\end{equation}
The explicit form of the densities can be found in the Appendix A.
As shown in Fig.\,\ref{fig.2}, there are eight fermionic bound
states in the potential well with $a=5$ and $V_0=1.2$.
In Fig.\,\ref{fig.4} we show the charge density of these bound states, along with their sum. 
The individual densities are depicted with various dashed lines and their energies are indicated on the graph. 
Note that each curve shows the charge density for two bound states with the same
energy but opposite signs and this is due to the symmetry in the fermionic spectrum. 
Solid line in this figure shows the total charge density of all bound states. 
Since each of the bound states is normalized, the area under the graph of total
bound states is eight. 
When $a$ is large, the densities are contained for the most part inside the region bounded by $\pm a$.
\par
\begin{center}
\begin{figure}[th]\hspace{1.3cm} \includegraphics[width=13.cm]{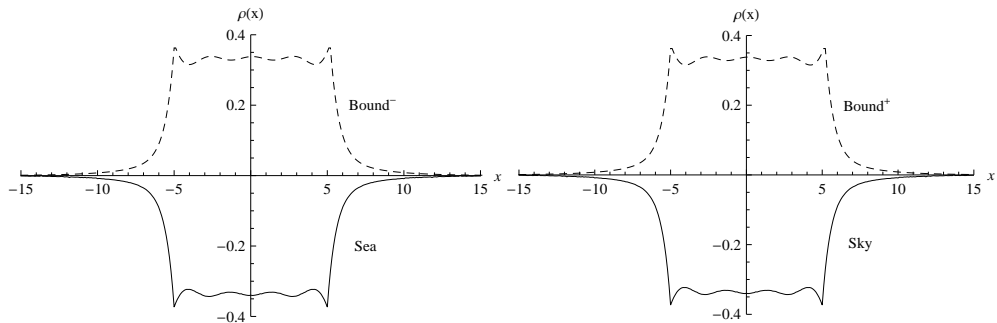}\caption{\label{fig.51} \small
Left graph: The spatial densities of the negative continuum
(Dirac sea) and the bound states with negative energy, and right
graph: The spatial densities of the positive continuum (Dirac sky)
and the bound states with positive energy for the potential well
with $a=5$ and $V_0=1.2$.}
\end{figure}
\end{center}

 
Now, we compute the total changes in the densities of the negative 
($\rho^{<}(x)$) and positive ($\rho^{>}(x)$) parts of the spectrum separately, and find that these changes are zero. 
These densities are defined by
 $\rho^{<}(x)=\rho_{\text{sea}}(x)+\rho^{<}_{\text{bound}}(x)$ and 
$\rho^{>}(x)=\rho_{\text{sky}}(x)+\rho^{>}_{\text{bound}}(x)$, where $\rho^{{\begin{smallmatrix} >\vspace{-.08cm}\\ <
\end{smallmatrix}}}_{\text{bound}}(x)$ denote the densities of all the bound states with positive and negative energy.
The left graph in Fig.\,\ref{fig.51} displays $\rho_{\text{sea}}(x)$ and $\rho^{<}_{\text{bound}}(x)$ and the right graph of this figure shows $\rho_{\text{sky}}(x)$ and $\rho^{>}_{\text{bound}}(x)$ for the parameters $a=5$ and $V_0=1.2$. 
From these graphs, it is obvious that $\rho^{<}(x)$ and $\rho^{>}(x)$ are zero, separately. 
Therefore, the change in the total density of states including the spatial densities of the spectral deficiency in the negative continuum, positive continuum and the sum of bound states is identically zero at each point of space, i.e.\;it is exactly equal to the free case. 
This shows that the total number of states and the total density remain unchanged as
compared with the free case, and the total density remains uniform.
\section{The Casimir energy for this model}
In order to obtain the Casimir energy, we should subtract the
zero-point energy in the absence from the presence of the
 background field. 
We have already obtained an expression for the Casimir energy in Ref.\,\cite{la},
which can be expressed as follows
\begin{align}\label{e17a}\vspace{.2cm}
E_{\text{Casimir}}&=\left\langle\Omega\left|H\right|\Omega
\right\rangle-\left\langle0\left|H_{\text{free}}\right|0
\right\rangle\nonumber\\
&=\int_{-\infty}^{+\infty}\text{d}x\int_{0}^{+\infty}\frac{\text{d}p}{2\pi}\sum\limits_{j=\pm}\left(-\sqrt{p^2+m_0^2}\right)\nu_p^{j\dag}\nu_p^j+
\int_{-\infty}^{+\infty}\text{d}x\sum\limits_{i}\left(E_{\text{bound}}^{i<}\right)\chi_{2\text{b}_i}^\dag\chi_{2\text{b}_i}\nonumber\\
&-\int_{-\infty}^{+\infty}\text{d}x\int_{-\infty}^{+\infty}\frac{\text{d}k}{2\pi}\left(-\sqrt{k^2+m_0^2}\right)v_k^\dag v_k,
\end{align}
where the minus superscript on $E^{i<}_{\text{bound}}$ denotes the bound state with negative energy, and we have denoted the vacuum state in the presence (absence) of the background field by $|\Omega\rangle$ ($|\,0\rangle$). 
Note that for our problem the whole spectrum is symmetric with respect to the line of $E=0$. 
Therefore, the expression for the Casimir energy, which is only in terms of negative energies, is obviously equivalent to the conventional one where one would sum over all modes symmetrically with a factor of $1/2$, while preserving the sign.
It is worth noticing that we have seen that even for the models without such a symmetry in the spectrum, like the one in \cite{la}, the same argument is true and as we stated there, we can calculate the Casimir energy using only the negative-energy states, or only the positive-energy states, or the sum of all the states divided by two and these three ways are equivalent.
\par
\begin{center}
\begin{figure}[th] \hspace{1.3cm}\includegraphics[width=13.cm]{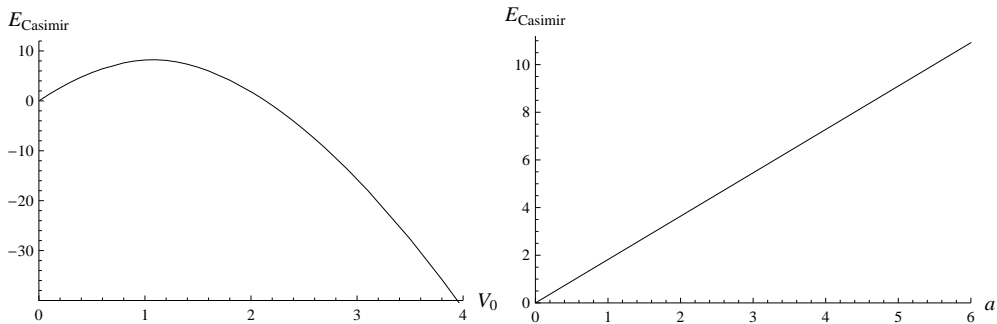}\caption{\label{fig.7} \small
Left graph: The Casimir energy as a function of $V_0$ at $a=5$.
Right graph: The Casimir energy as a function of $a$ at $V_0=1.2$.}
\end{figure}
\end{center}

 
Substituting the expressions for the eigenstates in the absence and presence of the potential into Eq.\,(\ref{e17a}), we obtain the Casimir energy for our model. 
In the left graph of Fig.\,\ref{fig.7} we show the Casimir energy as a function of $V_0$ for $a=5$. 
As can be seen, there is a maximum in this graph which occurs when the bound states change the direction and start to return to their continuum of origin. 
At $V_0=2$, the Casimir energy becomes zero and when the depth of the potential well is increased from this value, the Casimir energy becomes negative.
Note that, as can be seen in Fig.\,\ref{fig.2}, after $V_0=2$ there remain only the first two bound levels, with their energies approaching zero. 
Therefore, for these values of the depth, the only effect of the well on the spectrum 
is the change in the energy levels of the continua and the 
Casimir energy for $V_0>2$ comes from these changes. 
\par 
We also show the Casimir energy as a function of the width of the potential well, $a$, for $V_0=1.2$ in the right graph of Fig.\,\ref{fig.7}. 
As can be seen, the Casimir energy is a linearly increasing function of $a$ for fixed $V_0$.
Since regardless of the value of $V_0$ none of the bound states crosses $E=0$ (see Fig.\,\ref{fig.2}), there are no cusps in the graphs of the Casimir energy 
(see the cusps of the Casimir energy graphs in \cite{la,dehghan}).
\par 
We have observed that in the process in which the depth of the potential well increases from zero to its final value, 
spectral deficiencies develop in both of the continua and bound states appear. 
When the bound states return to the continua of origin, spectral deficiencies decrease. We compute the changes in the negative- and positive-energy densities of the spectrum, due to the aforementioned spectral deficiencies and the bound states. 
The changes in the energy densities of the Dirac sea and sky (i.e.\;the difference between the energy densities in the presence and absence of the background field) are given by
\begin{align}\label{e19}
\varepsilon_{\begin{smallmatrix} \text{sky}\\ \text{sea}
\end{smallmatrix}}(x)=\sum_{j=\pm}\int_{0}^{+\infty}\frac{\text{d}p}{2\pi}\; E \; \xi^{j\dag}_{p_{\begin{smallmatrix} \text{sky}\\ \text{sea}
\end{smallmatrix}}}(x)\; \xi^{j}_{p_{\begin{smallmatrix} \text{sky}\\ \text{sea}
\end{smallmatrix}}}(x)-
\int_{-\infty}^{+\infty}\frac{\text{d}k}{2\pi}\; E_{\text{free}}\;
\xi^{\text{free}\dag}_{k}(x)\;\xi^{\text{free}}_{k}(x),
\end{align}
where $E=\pm\sqrt{p^2+m_0^2}$, $E_{\text{free}}=\pm\sqrt{k^2+m_0^2}$
($\pm$ signs denote the positive and negative parities, respectively),
 $\xi_{p}^{j}=\mu^j_p\,(\nu_p^j)$ for the interacting Dirac sky (sea), and $\xi_k^{\text{free}}=u_k\,(v_k)$
 for the free Dirac sky (sea).
 The bound state energy densities are
\begin{equation}\label{e20}
\varepsilon_{\text{bound}}^{{\begin{smallmatrix} >\vspace{-0.08cm}\\ <
\end{smallmatrix}}}(x)=\sum_i E_{\text{bound}}^{i{\begin{smallmatrix} >\vspace{-0.08cm}\\ <
\end{smallmatrix}}}
\;\xi_{\text{b},i}^{\dag}(x)\;\xi_{\text{b},i}(x),
\end{equation}
where the ${\begin{smallmatrix} >\vspace{-0.08cm}\\ <
\end{smallmatrix}}$ superscripts refer to the sign of the bound state energies. 
The explicit form of the energy densities can be easily obtained using the expressions given in the Appendix A for the densities of the states.
\par
\begin{center}
\begin{figure}[th] \hspace{1.cm}\includegraphics[width=13.cm]{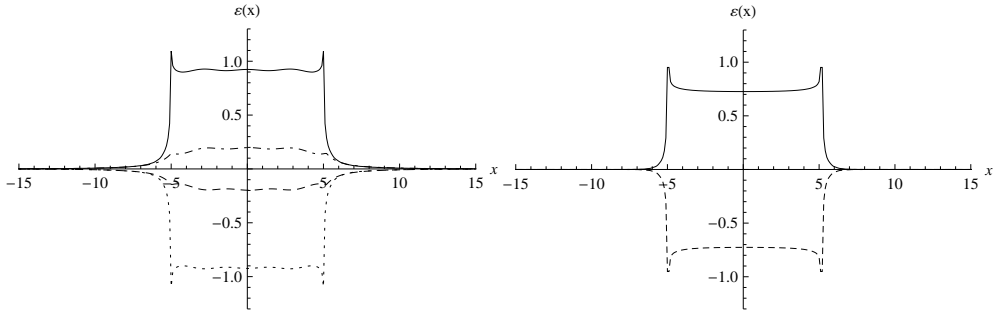}\caption{\label{fig.8} \small
The energy densities as a function of $x$ for a potential of the parameters $a=5$ and $V_0=1.2$. Left graph: Solid (dotted) line shows the energy density of the negative (positive) continuum states and dashed (dotdashed) line shows the sum of the energy densities of negative (positive) bound states. 
Right graph: Solid line shows the sum of the energy densities of the negative-energy bound and continuum states (the Casimir energy density) 
and dashed line shows the sum of
the energy densities of the positive-energy bound and continuum states.}
\end{figure}
\end{center}

 
We display all these energy densities for a potential well with $a=5$ and $V_0=1.2$ in Fig.\,\ref{fig.8}. 
The left graph in this figure shows the energy densities of the sum of bound states with negative ($\varepsilon_{\text{bound}}^{<}(x)$) and positive ($\varepsilon_{\text{bound}}^{>}(x)$) energy, separately. 
Moreover, this graph shows the energy densities of continuum states with negative ($\varepsilon_{\text{sea}}(x)$) and positive ($\varepsilon_{\text{sky}}(x)$) energies.
The sum of $\varepsilon_{\text{sea}}(x)$ and $\varepsilon_{\text{bound}}^{<}(x)$ which is the total energy density of states with negative energy has been depicted in the right graph of this figure. 
Note that this density is in fact the Casimir energy density ($\varepsilon_{\text{Casimir}}(x)$). 
Furthermore, we show the total energy density of states with positive energy
($\varepsilon^{>}(x)=\varepsilon_{\text{sky}}(x)+\varepsilon_{\text{bound}}^>(x)$),
 for comparison. 
As can be seen from the figure, $\varepsilon^{<}(x)$ and $\varepsilon^{>}(x)$ are exactly the mirror images of each other. 
This result shows that for the system chosen here, as we stated before, the Casimir energy can be calculated by any of the following relations
\begin{align}\label{r3b1}\vspace{.2cm}
 E_\text{Casimir}&=\int_{-\infty}^{+\infty}\text{d}x\;\varepsilon_{\text{sea}}(x)+\sum_iE_{\text{bound}}^{i<}
 =-\left(\int_{-\infty}^{+\infty}\text{d}x\;\varepsilon_{\text{sky}}(x)+\sum_iE_{\text{bound}}^{i>}\right)\nonumber\\
 &=\frac{1}{2}\left(\int_{-\infty}^{+\infty}\text{d}x\;\varepsilon_{\text{sea}}(x)+\sum_iE_{\text{bound}}^{i<}\right)-\frac{1}{2}\left(\int_{-\infty}^{+\infty}\text{d}x\;\varepsilon_{\text{sky}}(x)+\sum_iE_{\text{bound}}^{i>}\right).
 \end{align}
\section{Stability of the solutions}
\begin{center}
\begin{figure}[th] \hspace{1.3cm}\includegraphics[width=13.cm]{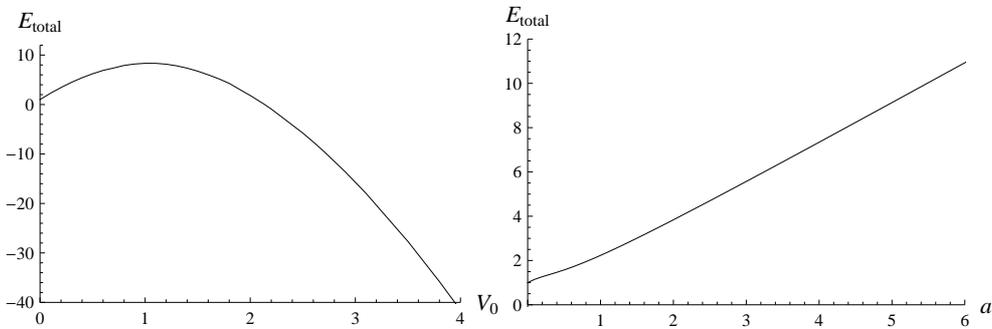}\caption{\label{stability} \small
The graphical representation of the total energy (the sum of the energy of a valence fermion in the ground state and the Casimir energy).  
Left graph shows the total energy as a function of $V_0$ when $a=5$ and right graph shows this energy as a function of $a$ when $V_0=1.2$.}
\end{figure}
\end{center}

In this section we consider a system consisting of a single fermion present in the lowest fermionic bound state, taking into account the Casimir energy. 
The total energy for such a system is the sum of the energy of the first fermionic bound state and the Casimir energy. 
Figure \ref{stability} shows this energy. 
The left graph shows the total energy as a function of the depth of the potential, $V_0$, when $a=5$ and the right graph shows this energy as a function of the width of the potential, $a$, when $V_0=1.2$. 
As can be seen, there is no minimum in these two graphs. 
From these graphs we conclude that since the total energy exceeds the fermion mass for most choices of the parameters, the first bound state is unstable, except for a very small region of the depth of the potential near $V_0=2$. 
This result differs from that of the system in \cite{dehghan}, where the potential is chosen to be an electric potential. 
Since the order of the Casimir energy is in general larger than the fermion mass, for most choices of the parameters of the potential the main portion in the total energy comes from the Casimir energy and this is the reason why the graphs of the total energy are so similar to the graphs of the Casimir energy.
\section{Conclusion}
In this paper we have computed the vacuum polarization and Casimir energy for a very simple model. 
The model includes a Fermi field coupled to a scalar potential in ($1+1$) dimensions. Since the scalar potential has the simple form of a symmetrical square-well,
we are able to obtain the whole spectrum of the coupled Dirac field. 
This model possesses all the symmetries C, P, and T, separately. 
In this model the charge conjugation operator relates the positive-energy solutions of the fermion to the negative-energy ones.
Due to this symmetry, the energy spectrum of the system is completely symmetric, 
i.e.\;for every positive-energy solution there is a solution with an energy of the same absolute value but opposite sign. 
This symmetry is obvious in the graph of the bound state energies. We have computed the spectral deficiency in the continua and found that as we increase the depth of the potential with the appearance of the bound states, deficiencies develop in both continua and when the bound states re-join the continua, deficiencies decrease. 
We also observe that the general trend for deficiencies in both of the continua is
exactly the same, when the potential depth increases from zero.
Moreover, we have displayed the densities of bound states and densities of deficiencies in the continua. 
We have concluded that due to the symmetries of the model not only the total density
remains unchanged as compared with the free case, but this happens also for the total negative and positive densities, separately. 
The result that the total number of states and the total density of states remain unchanged as compared with the free case, confirms that the spectrum remains complete in the presence of the potential well.
The vacuum polarization has been computed for this model and as is obvious from the completeness of negative states, the VP is zero for any choice of the parameters of the potential well. 
In the second part of the paper we have computed the Casimir energy for our model. Since we have all the eigenfunctions and eigenvalues of the system, we are able to calculate the Casimir energy of the system by the direct subtraction of the zero-point energy in
the presence and absence of the disturbance. 
The interesting result is that although the vacuum polarization is always zero for the square-well potential, due to the charge conjugation symmetry, the Casimir energy is not in general zero. 
In the graph of the Casimir energy as a function of the depth of the potential there is a maximum which occurs when the negative-energy bound levels change direction and start to return to the continua of their origin. 
When the potential depth is $2(m_0)$, the Casimir energy is zero and after this depth the Casimir energy is always negative. 
We have also displayed the Casimir energy as a function of the width of the well and found that the Casimir energy increases linearly as the width of the well increases and since none of the bound energy levels crosses the line of $E=0$, there is no cusp in this graph. 
Then, we have depicted the Casimir energy density and the density of all the states with positive energy for comparison and found that these two densities are exactly the mirror images of each other. 
Finally, considering a system consisting of a valence fermion present in the ground state, we conclude that the lowest fermionic bound state is almost always unstable for different choices of the parameters of the scalar potential.

\appendix
\section{} \label{Calculation}
\setcounter{equation}{0}
\renewcommand{\theequation}{\Alph{section}.\arabic{equation}}
The explicit expressions for the coefficients of the bound states
given in Eq.\,(\ref{e5}) are as follows
\begin{align}\label{1}\vspace{.2cm}
&N=\text{Normalization of bound states},\nonumber\\
&b=N\left[\frac{\lambda(m_0-V_0)-iEV_0}{m_0\mu}\sin[2a\mu]+\cos[2a\mu]\right]\nonumber,\\
&c=\frac{-N}{2m_0\mu}\text{e}^{i\mu
a}\left[EV_0-m_0\mu+i\lambda(m_0-V_0)\right],\nonumber\\
&d=\frac{N}{2m_0\mu}\text{e}^{-i\mu
a}\left[EV_0+m_0\mu+i\lambda(m_0-V_0)\right],
\end{align}
where the normalization factor for the bound states is
\begin{align}\label{2}\vspace{.2cm}
N&=\bigg\{\frac{1}{\lambda}-\frac{\lambda}{m_0(m_0-V_0)}
+\sin\left[4a\text{Re}(\mu)\right]\nonumber\\&\times\bigg[\frac{1}{4\text{Re}(\mu)}\left(2-\frac{\lambda^2\left(E^2+(m_0-V_0\right)^2)+E^2V_0^2}{m_0^2|\mu|^2}
-\frac{m_0^2|\mu|^4-2m_0V_0E^2|\mu|^2+E^4V_0^2}{m_0^2|\mu|^2(m_0-V_0)^2}\right)
\nonumber\\&+\frac{(m_0-V_0)\text{Re}(\mu)}{m_0|\mu|^2}\bigg]\nonumber\\&+\cos\left[4a\text{Re}(\mu)\right]\left[\frac{-\lambda
E^2+\lambda|\mu|^2-\lambda(m_0-V_0)^2}{2m_0|\mu|^2(m_0-V_0)}-
\frac{\lambda^2(m_0-2V_0)-m_0V_0^2}{2m_0\lambda|\mu|^2}+\frac{1}{2\lambda}\right]
\nonumber\\&+\sinh\left[4a\text{Im}(\mu)\right]\nonumber\\&\times\bigg[\frac{1}{4\text{Im}(\mu)}
\left(2+\frac{\lambda^2(E^2+(m_0-V_0)^2)+E^2V_0^2}{m_0^2|\mu|^2}+\frac{m_0^2|\mu|^4-2m_0V_0E^2|\mu|^2+E^4V_0^2}{m_0^2|\mu|^2(m_0-V_0)^2}\right)
\nonumber\\&+\frac{(m_0-V_0)\text{Im}(\mu)}{m_0|\mu|^2}\bigg]\nonumber\\&+\cosh\left[4a\text{Im}(\mu)\right]\left[\frac{\lambda|\mu|^2+\lambda
E^2+\lambda(m_0-V_0)^2}{2m_0|\mu|^2(m_0-V_0)})+\frac{m_0V_0^2
+\lambda^2(m_0-2V_0)}{2
m_0\lambda|\mu|^2}+\frac{1}{2\lambda}\right]\bigg\}^{-1/2}.
\end{align}
The explicit expressions for the coefficients of the continuum
states given in Eq.\,(\ref{e7}) are as follows
\begin{align}\label{3}\vspace{.2cm}
&N_{\pm}=\text{Normalization of continuum states},\nonumber\\
&h=\frac{N_{\pm}}{2p}\left(\text{e}^{-i\mu
a}[\frac{m_0(m_0-V_0)}{(E-\mu)}+(p-E)]+\text{e}^{i\mu a}[\pm m_0\pm
\frac{(V_0-m_0)(E-p)}{(E-\mu)}]\right)\nonumber,\\
&k=\frac{N_{\pm}}{2p}\left(\text{e}^{-i\mu
a}[\frac{m_0(V_0-m_0)}{(E-\mu)}+(p+E)]+\text{e}^{i\mu a}[\mp m_0\mp
\frac{(V_0-m_0)(E+p)}{(E-\mu)}]\right),
\end{align}
where the normalization of the continuum states is ($\pm$ signs refer to the parity of the states)
\begin{align}\label{4}\vspace{.2cm}
N_{\pm}&=\bigg\{\cosh\left[2a\text{Im}(\mu)\right]\left[\frac{2E^2(V_0^2-m_0^2)}{p^2\left(E^2+|\mu|^2-2E\text{Re}(\mu)\right)}+\frac{2E^2}{p^2}-\frac{4m_0E(V_0-m_0)\text{Re}(\mu)}{p^2\left(E^2+|\mu|^2-2E\text{Re}(\mu)\right)}\right]
\nonumber\\&\mp2\cos\left[2a\text{Re}(\mu)\right]\left[\frac{m_0E}{p^2}+\frac{m_0E(V_0-m_0)^2+2E^2(V_0-m_0)(E-\text{Re}(\mu))}{p^2\left(E^2+|\mu|^2-2E\text{Re}(\mu)\right)}\right]\bigg\}^{-1/2}.
\end{align}
The change in the number of levels of the
continua are as follows
\begin{align}\label{5}\vspace{.2cm}
Q_{\begin{smallmatrix} \text{sky}\\ \text{sea}
\end{smallmatrix}}&=-\frac{1}{2}+\int_{0}^{+\infty}\frac{\text{d}p}{\pi
 p^2}\bigg\{-2ap^2\nonumber\vspace{0.3cm}\\&+(|N_{+}|^2+|N_{-}|^2)\sinh[2a\text{Im}(\mu)]\nonumber\vspace{0.2cm}\\&\times\left[\frac{2m_0(m_0-V_0)
 \text{Im}(\mu)}{E^2+|\mu|^2-2E\text{Re}(\mu)}+\frac{p^2}{\text{Im}(\mu)}
 \left(1+\frac{(m_0-V_0)^2}{E^2+|\mu|^2-2E\text{Re}(\mu)}\right)\right]\nonumber\\&+(|N_{+}|^2-|N_{-}|^2)\sin[2a\text{Re}(\mu)]
 \nonumber\\&\times\left[m_0\left(1-\frac{(m_0-V_0)^2}{E^2+|\mu|^2-2E\text{Re}(\mu)}\right)+\frac{2p^2(m_0-V_0)(E-\text{Re}(\mu))}{\text{Re}(\mu)(E^2+|\mu|^2-2E\text{Re}(\mu))}\right]\bigg\}.
\end{align}
The spatial charge densities for the bound states and the continua
are obtained in the following forms
\par for $|x|\geqslant a$:
\begin{align}\label{6}\vspace{.2cm}
\rho_{\begin{smallmatrix} \text{sky}\\ \text{sea}
\end{smallmatrix}}(x)&=\int_{0}^{\infty}\frac{\text{d}p}{\pi
p^2}\bigg\{(|N_{+}|^2+|N_{-}|^2)\bigg[\cosh\left[2a\text{Im}(\mu)\right]\cos\left[2p(|x|-a)\right]\nonumber\\&
\times\left(\frac{2m_0E(m_0-V_0)(E-\text{Re}(\mu))}{E^2+|\mu|^2-2E\text{Re}(\mu)}-m_0^2-\frac{m_0^2(m_0-V_0)^2}{E^2+|\mu|^2-2E\text{Re}(\mu)}\right)
\nonumber\\&+\sinh\left[2a\text{Im}(\mu)\right]\sin\left[2p(|x|-a)\right]\left(\frac{2m_0p(m_0-V_0)\text{Im}(\mu)}{E^2+|\mu|^2-2E\text{Re}(\mu)}\right)\bigg]
\nonumber\\&+(|N_{+}|^2-|N_{-}|^2)\bigg[\cos\left[2a\text{Re}(\mu)\right]\cos\left[2p(|x|-a)\right]\nonumber\\&\times
\left(m_0E+\frac{m_0E(m_0-V_0)^2}{E^2+|\mu|^2-2E\text{Re}(\mu)}-
\frac{2m_0^2(m_0-V_0)(E-\text{Re}(\mu))}{E^2+|\mu|^2-2E\text{Re}(\mu)}\right)
\nonumber\\&+\sin\left[2a\text{Re}(\mu)\right]\sin\left[2p(|x|-a)\right]
\left(m_0p-\frac{m_0p(m_0-V_0)^2}{E^2+|\mu|^2-2E\text{Re}(\mu)}\right)\bigg]\bigg\},\nonumber\\
\rho_{\text{bound}}^{\pm}(x)=& 2N^2
\text{e}^{-2\lambda\left(|x|-a\right)},
\end{align}
\par and for $|x|\leqslant a$:
\begin{align}\label{7}\vspace{.2cm}
\rho_{\begin{smallmatrix} \text{sky}\\ \text{sea}
\end{smallmatrix}}(x)=&\int_{0}^{+\infty}\frac{\text{d}p}{\pi
}\bigg\{ -1
+\left(|N_{+}|^2+|N_{-}|^2\right)\left[1+\frac{(m_0-V_0)^2}{E^2+|\mu|^2-2E\text{Re}(\mu)}\right]
\cosh\left[2\text{Im}\left(\mu\right)x\right]\nonumber\\
&+2(m_0-V_0)\left(|N_{+}|^2-|N_{-}|^2\right)\left[\frac{E-\text{Re}(\mu)}{E^2+|\mu|^2-2E\text{Re}(\mu)}\right]
\cos\left[2\text{Re}\left(\mu\right)x\right]
\bigg\},\nonumber\\
\rho_{\text{bound}}^{\pm}(x)&=\frac{ N^2}{2m_0^2|\mu|^2}
\bigg\{\cosh\left[2\text{Im}(\mu)(x+a)\right]\nonumber\\&\times\bigg[\left(\lambda^2+\frac{E^2V_0^2+m_0^2|\mu|^2}{(m_0-V_0)^2}\right)
(E^2+|\mu|^2)-\frac{4m_0V_0E^2\text{Re}(\mu)^2}{(m_0-V_0)^2}+\lambda^2(m_0-V_0)^2\nonumber\\&+E^2V_0^2+m_0^2|\mu|^2\bigg]
+2\sinh\left[2\text{Im}(\mu)(x+a)\right]m_0\lambda
\text{Im}(\mu)\left[\frac{E^2+|\mu|^2}{(m_0-V_0)}+m_0-V_0\right]
\nonumber\\&+\cos\left[2\text{Re}(\mu)(x+a)\right]\nonumber\\&\times\bigg[\left(\lambda^2+\frac{E^2V_0^2-m_0^2|\mu|^2}{(m_0-V_0)^2}\right)
(|\mu|^2-E^2)-\frac{4m_0V_0E^2\text{Im}(\mu)^2}{(m_0-V_0)^2}-\lambda^2(m_0-V_0)^2\nonumber\\&-E^2V_0^2+m_0^2|\mu|^2\bigg]+\sin\left[2\text{Re}(\mu)(x+a)\right]m_0\lambda
\text{Re}(\mu)\left[\frac{E^2-|\mu|^2}{(m_0-V_0)}+m_0-V_0\right]\bigg\}.
\end{align}

 \end{document}